# GENERALIZING A HIGHLY CONFIGURABLE ANALYTICS PIPELINE TO REPLICATE AND SUPPORT EDUCATIONAL RESEARCH ACROSS MULTIPLE DOMAINS

**Y. Bai, P. Thajchayapong, A. Goel**

*Georgia Institute of Technology (UNITED STATES)*

## Abstract

Artificial intelligence assistants deployed in online learning environments create new opportunities to collect large volumes of learner interaction data and generate insights to improve student outcomes. Architecture for AI-Augmented Learning (A4L) is a modular data architecture that enables the collection, integration, and analysis of learner interaction data from educational AI systems, supporting the generation of instructional insights that facilitate personalized learning and reinforce the bidirectional feedback loop between instructors and learners. This study examines the modular design of the A4L Data Analytics Pipeline, an extensible data infrastructure that enables the ingestion, processing, and analysis of heterogeneous datasets generated by educational AI assistants. We describe the design principles and development process used to extend the pipeline's analytical capabilities while preserving generalizability across domains. We evaluate the pipeline through case studies spanning three research domains corresponding to three educational AI assistants deployed in online learning environments at Georgia Tech. Results show that a common set of statistical analysis methods can be consistently applied across datasets with differing structures and instructional contexts, enabling the pipeline to reproduce key analytical findings across domains. Furthermore, we demonstrate how analytical capabilities initially developed for one domain can be extended to support richer analyses in another, illustrating the pipeline's modularity and extensibility. These findings suggest that the A4L Analytics Pipeline can serve as reusable infrastructure for analyzing data generated by future educational AI assistants. By enabling analytics that can be systematically extended to new domains, the pipeline provides a foundation for deriving insights that inform the design, deployment, and evaluation of educational AI systems.

## 1 INTRODUCTION

Learning analytics is the analysis of data about learners to better understand and optimize how learning occurs, providing data-driven insights into the topic [1]. Educational inventions, such as online classrooms, Learning Management Systems (LMS), and educational AI assistants provide new ways to capture data for analysis. By leveraging these emerging sources of data, learning analytics may provide insights into which students need additional academic support to increase to their chances of success in class [2]. Learning analytics platforms also have the potential to analyse data from multiple sources to help institutions create unique learning opportunities to match each students' individual needs [3].

Total Learning Architecture [4], Datashop [5], Unizin [6], and My Learning Analytics [7] are examples of learning analytics systems that standardize and store data for educational purposes. A learning analytics system developed by the University of Phoenix [8] used a repository containing data from across the organization to facilitate analytics on predicting student persistence in their program using available data like grades, content usage, and demographics.

Researchers have developed data analytics pipelines as part of educational data architectures, such as a pipeline for classifying engagement behaviours in online learning environments [9] and a pipeline for converting educational systems data into knowledge components [10]. Although the research on these pipelines mentions generalizability, this topic is not the research's focus and deserves to be explored in greater detail. Research in learning analytics systems has found that extensibility challenges deserve more attention, and that real-world implementations of extensible and flexible analytics platforms are scarce [11]. In response to the extensibility challenge, one research group demonstrated that a learning analytics platform would benefit from a modular architecture that supports the addition of new analytics methods [12]. The real-world implementation portion remains an area for further investigation.

Architecture for AI-Augmented Learning (A4L) has developed the A4L Analytics Pipeline which provides evidence of a real-world implementation of a generalizable analytics pipeline for learning analytics. A4L is a modular data architecture that enables the collection, integration, and analysis of learner interaction data from educational AI systems, supporting the generation of instructional insights that facilitate personalized learning and reinforce the bidirectional feedback loop between instructors and learners [13]. The A4L Analytics Pipeline is one component of the overall A4L pipeline that enables the continuous flow of data from sources such as LMS and educational AI assistants [14]. The A4L Analytics Pipeline is designed as a Highly Configurable System (HCS) that supports flexible data analysis based on user input at runtime [15]. Santana et al 2025 [16] showed the use of HCS principles in the Analytics Pipeline results in a pipeline that is extensible and flexible. They found that the pipeline strikes a healthy balance between flexibility and structural coherence; structural coherence is defined as the absence of invalid system configurations.

In this research, we improve features of the existing A4L Analytics Pipeline by increasing the pipeline's analysis capabilities and making those capabilities applicable to a broad range of datasets. Then we test this pipeline by replicating real-world analyses from existing research on educational AI tools deployed in different classes of Georgia Tech's online program in computer science [17].

# 2 ANALYTICS PIPELINE ARCHITECTURE

## 2.1 Context in A4L Environment

The A4L Analytics Pipeline is an extendible and flexible platform for reproducible educational data analysis. The A4L Data Engine is the component of A4L that securely processes and organizes data from a wide variety of educational systems so that the data may be used for analytics and personalization of learning. The A4L Data Engine loads processed data into data stores, including a published data store for anonymized data. The Analytics Pipeline retrieves data from the published data store for analysis. Another component of A4L is the A4L Visualization Pipeline which provides insight into the educational data ingested by A4L through role-specific dashboards for researchers, instructors, and learners. The Analytics Pipeline generates analysis results that the Visualization Pipeline displays to researchers and instructors in their dashboards, as shown in figure 1 [14].

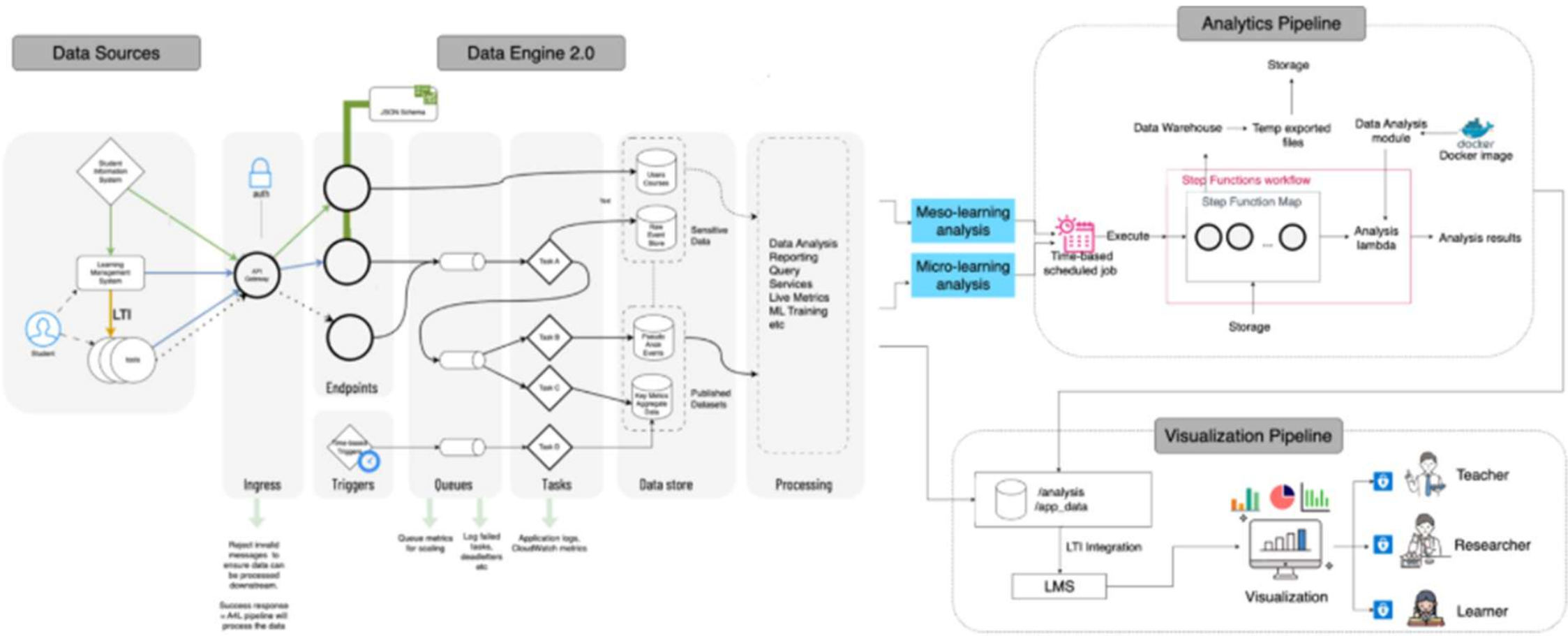

*Figure 1. A4L (Architecture for AI-Augmented Learning) components in the A4L Pipeline's overall conceptual architecture [14]. The A4L Data Engine securely processes educational system data. The A4L Analytics Pipeline performs statistical analyses on the processed data. The A4L Visualization Pipeline presents analysis results from the Analytics Pipeline to researchers and instructors through dashboards.*

## 2.2 Analytics Pipeline Components

The A4L Analytics Pipeline is defined as a state machine that runs in a cloud environment. The state machine takes a JSON-formatted analysis configuration payload as input. This payload contains instructions for how the pipeline should run.

A run of the Analytics Pipeline begins with the data fetch module, which retrieves datasets specified in the payload by querying the data warehouse and stores the data in a staging area. After all datasets have been loaded in the staging area, the data fetch module passes the locations of each dataset file and the payload to the analysis module.

The analysis module accesses the data saved in the staging area and, according to the instructions in the payload, performs pre-processing, data transforms, and data analysis on the retrieved data. Analysis results are uploaded to the cloud environment's durable object storage in a specific directory according to the payload. The Visualization Pipeline accesses the analysis results from this storage location. Much of this research focused on modifying the analysis module to perform more data processing and analytics procedures.

In the overall A4L environment, the Analytics Pipeline runs daily to perform analytics on new data available in the datastore. These daily scheduled runs allow the pipeline to automatically generate up to date analysis results, ensuring data visualizations shown to instructors and researchers are current.

This research added components to the Analytics Pipeline to support the pipeline's generation of analysis results in a timely manner. The daily scheduled job was modified to monitor the published data store for new data ingested by the data engine. If new data are detected, the scheduled job updates the data warehouse with new data and archives any old data that would be overwritten. The scheduled job identifies which payloads should be run through the Analytics Pipeline by determining if a payload uses any datasets that were updated in this process. Figure 2 illustrates the orchestration of the pipeline's components.

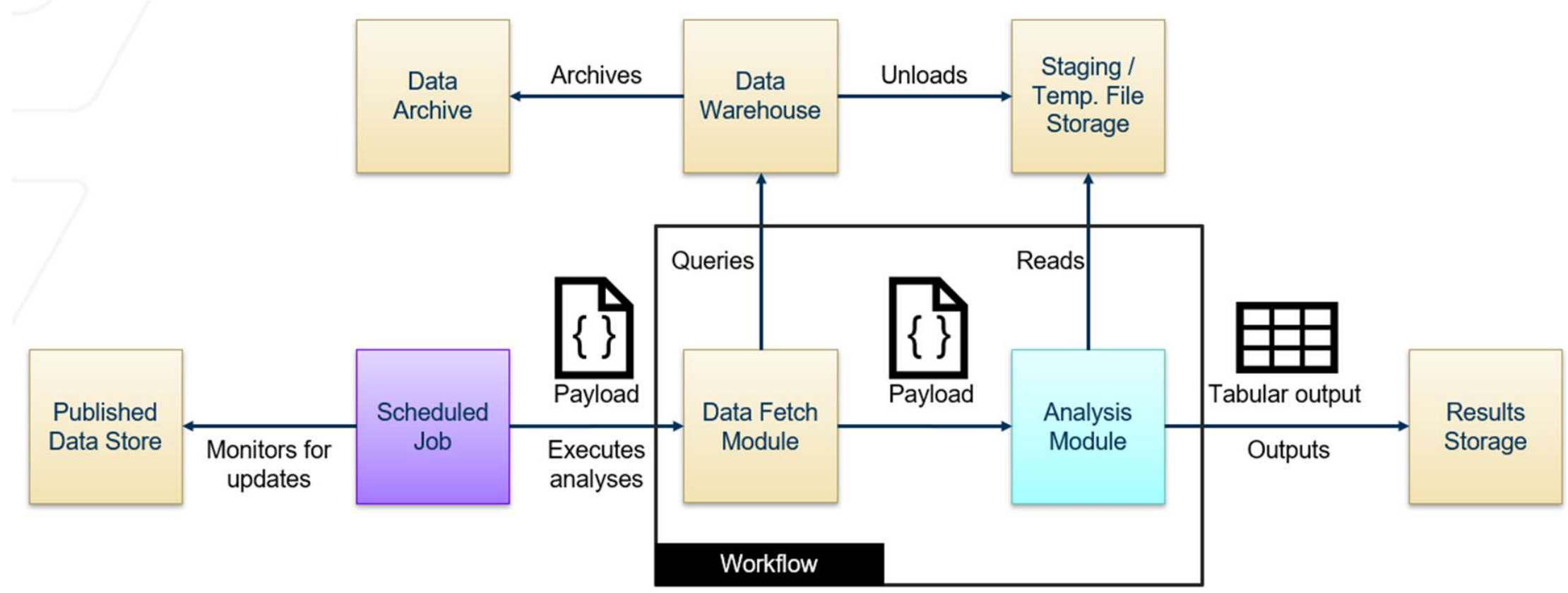

*Figure 2. Orchestration of A4L Analytics Pipeline components, adapted from Santana et al 2025 [16].*

## 3 EDUCATIONAL AI ASSISTANT RESEARCH DOMAINS

Research conducted on the following educational AI assistants was replicated using the Analytics Pipeline. The research domain associated with each AI assistant is described in the sections that follow.

### 3.1 Jill Watson

Jill Watson (JW) is a virtual teaching assistant that helps answer students' questions in the classroom's online discussion forum [18]. A4L collected data related to the deployment of JW in a graduate level computer science class at Georgia Tech for the Fall 2023 and Spring 2024 semesters. These data include information from the course's LMS, student demographics, and chat logs with JW. Santana et al 2025 [16] analyzed these data to understand if demographic factors influence the adoption of AI technology and if adoption of the technology affects performance in the course. This analysis used statistics such as Welch's t-tests to compare students' final scores between JW users and non-users.

### 3.2 Virtual Ecological Research Assistant

The Virtual Ecological Research Assistant (VERA) is used to simulate ecological phenomena based on conceptual models [19]. A4L collected data related to the deployment of VERA in a graduate level computer science class Georgia Tech for the Summer 2023 semester. These data include student demographics, student actions using VERA, online forum posts, and psychometric survey data. Fryer et al 2025 [20] analyzed these data to discover that AI tool adopters had significantly higher need for cognition (NFC) scores than non-adopters. These findings were supported by evidence from statistical results such as Welch's t-tests and statistical power of t-tests applied to metrics between AI tool adopters and non-adopters.

### 3.3 Social Agent Mediated Interactions

Social Agent Mediated Interactions (SAMI) is an AI assistant the helps foster social interactions in online classrooms [21]. A4L collected data related to the deployment of SAMI in a graduate level computer science class at Georgia Tech for the Fall 2023, Spring 2024, Summer 2024, and Fall 2024 semesters. These data include student demographics, student interactions with SAMI, online forum posts, and psychometric survey data. From these data, Wójcik et al 2025 [22] found there was no bias in SAMI adoption, promoting an equitable environment for learners. This research also found that students who used SAMI had a higher sense of belonging (SOB) compared to non-users. Welch's t-tests and Mann-Whitney U tests between students' SOB scores for SAMI users and non-users provided evidence to support these findings.

## 4 METHODOLOGY

To support the statistical analyses used across the previously described research domains, the Analytics Pipeline was developed as a highly configurable system (HCS), enabling a single underlying codebase to be applied to different scenarios through flexible configuration. Statistical methods, including Welch's t-test, were applied across all three domains, making it essential for the pipeline's implementations of these tests to be sufficiently flexible to support cross-domain use without requiring code modification.

## 5 RESULTS

### 5.1 Implementing Generalizable Analysis Functionality

Using the implementation of Welch's t-test as an example, we first analyzed variations in how each research domain applied the t-test in their research. We noted that both one-sided and two-sided alternative hypotheses were used. Therefore, the Analytics Pipeline's t-test must allow for configurable alternative hypotheses. Each domain used a different tabular dataset for analysis, so the pipeline's t-test must allow the user to specify the dataset on which to perform analysis. Each domain used different columns from their respective tabular datasets as the independent and dependent variables of the t-test, so the pipeline's t-test must allow the user to specify which columns to use as the independent and dependent variables. Finally, the storage location for the analysis results is different for each domain. The pipeline must allow the user to specify the storage location in the cloud environment, helping the Visualization Pipeline understand which analysis results correspond to which AI assistant for display in their designated dashboard.

All the configurable options for the t-test were programmed as variables in the pipeline's codebase. The values of those variables are retrieved from the analysis configuration payload by getting the assigned value for each key present in the JSON file. Table 1 demonstrates how the keys of the analysis configuration JSON differ between the research domains. The JSON key values for each input differ for each research domain, but the same underlying code in the Analytics Pipeline's analysis module is called. Other statistical procedures were similarly implemented to extend the Analytics Pipeline's analysis capabilities in a flexible manner.

*Table 1. Analysis Configuration for performing Welch's t-test across three different research domains, JW, VERA, and SAMI. The variable key values are fed to the same code in the Analytics Pipeline's analysis module.*

| *Payload Component* | *JW Analysis Value* | *VERA Analysis Value* | *SAMI Analysis Value* |
|---|---|---|---|
| Alternative hypothesis | Two-sided | Two-sided | Less than |
| Dataset | Fall 2023 JW usage data | Summer 2023 VERA usage data | Fall 2024 SAMI usage data |
| Independent variable | Used-JW | Used-VERA | Used-SAMI |

| Dependent variable(s) | Course final score | NFC score, self-efficacy score, help-seeking score, peer-learning score | SOB score, distinct impressions score, comfortable interacting score, sense of collaboration score |
|---|---|---|---|
| Result file name | jw_fall23_ttest | vera_summer23_ttest | sami_fall24_ttest |

## 5.2 Generation of Additional Analysis Results

We demonstrate the results of extending the Analytics Pipeline in a flexible manner by applying statistical tests originating from one domain onto a different domain. Welch's t-test was used by each domain to determine what differences existed between AI tool users and non-users, however only the VERA domain applied a calculation of statistical power for Welch's t-test. This research applied the statistical power calculation to the SAMI domain by creating an analysis configuration payload that invoked the power function in the analysis module with parameters specific to the SAMI dataset. Table 2 compares how the input to the pipeline differs for calculating t-test power for VERA data from a computer science course in the summer of 2023 and for SAMI data from a different computer science course in the fall of 2024. The name of the desired statistic to perform is the only component that is equal between the two configurations. The other components of the payload are unique to the research domain.

*Table 2. Comparison of Analytics Pipeline payload configurations for calculating power for Welch's t-test for VERA in the summer of 2023 and SAMI in the fall of 2024*

| *Payload Component* | *VERA Analysis Value* | *SAMI Analysis Value* |
|---|---|---|
| Statistic name | get_welch_power | get_welch_power |
| Dataset | Summer 2023 VERA usage data | Fall 2024 SAMI usage data |
| Independent variable | Used-VERA | Used-SAMI |
| Dependent variable(s) | NFC score, self-efficacy score, help-seeking score, peer-learning score | SOB score, distinct impressions score, comfortable interacting score, sense of collaboration score |
| Result file name | vera_summer23_ttest_power | sami_fall24_ttest_power |

All analysis payload components in table 2 are user-specified. A user wanting to calculate power for Welch's t-test selects get_welch_power as the option for the statistic name. Then, the user is free to specify the other options based on the dataset in question. The results of using the Analytics Pipeline to calculate statistical power for Welch's t-test for the SAMI fall 2024 analysis are shown in table 3. The results produced by the pipeline complement Wójcik et al 2025's [22] findings by providing more quantitative information for the SAMI research domain.

*Table 3. Analytics Pipeline statistics for selected Community of Inquiry (CoI) survey questions from SAMI fall 2024 data. The t statistics and p-values are from Wójcik et al 2025's research [22]. The power values are from leveraging the Analytics Pipeline's extensibility and flexibility.*

| *Variable* | *t* | *p-value* | *Power* |
|---|---|---|---|
| Sense of Belonging | 1.41 | 0.080 | 0.403 |
| Comfortable Interacting | 1.87 | 0.032 | 0.642 |

## 5.3 Replication of Analysis Results

Configuration payloads were constructed for each research domain and passed as input to the Analytics Pipeline. The outputs generated by the pipeline replicate the findings of each domain's data analysis.

The result files were generated in the respective bucket for each AI tool, and these files are available for use by the Visualization Pipeline for display in dashboards dedicated to each AI tool.

For example, the Analytics Pipeline's contingency table option was applied to all three domains and produced descriptive statistic results that matched previous findings. For the JW domain, the contingency table used JW usage and student age group. For the VERA domain, the table used JW usage in VERA and student gender. For the SAMI domain, two tables were used: SAMI usage and student age and SAMI usage and student gender.

*Table 4. A subset of statistical options available in the Analytics Pipeline's analysis module and their usage across the JW, VERA, and SAMI research domains.*

| *Statistic* | *JW usage* | *VERA usage* | *SAMI usage* |
|---|---|---|---|
| Welch's t-test | Yes | Yes | Yes |
| Contingency table | Yes | Yes | Yes |
| Power of Welch's t-test | No | Yes | Not originally |

## 6 DISCUSSION

By implementing extensible and flexible statistical options in the Analytics Pipeline, results from research domains on different educational AI assistants could be replicated using one unified codebase. We have also demonstrated that this unified codebase is flexible and extensible enough to enhance data analysis by applying analytical techniques across research domains.

The results from this research suggest that the statistical capabilities of the Analytics Pipeline are generalizable and can work with data sources that may not currently exist. We use XYZ as a placeholder for an educational AI assistant that has yet to be deployed in a classroom. When data on XYZ's classroom usage and context are made available, a researcher will be able to create an analysis configuration payload and serve that as input to the Analytics Pipeline for analysis. Like how power of Welch's t-test was applied from VERA data analysis to SAMI data analysis, XYZ could make use of statistical tests originally used by other research domains and apply those tests to its data through the power of configurability of the Analytics Pipeline. No code will need to be written for XYZ to get started with using the analytics capabilities of the pipeline. Figure 3 illustrates a possible use case for a future AI assistant that takes statistical options from the Analytics Pipeline and applies them to its own datasets.

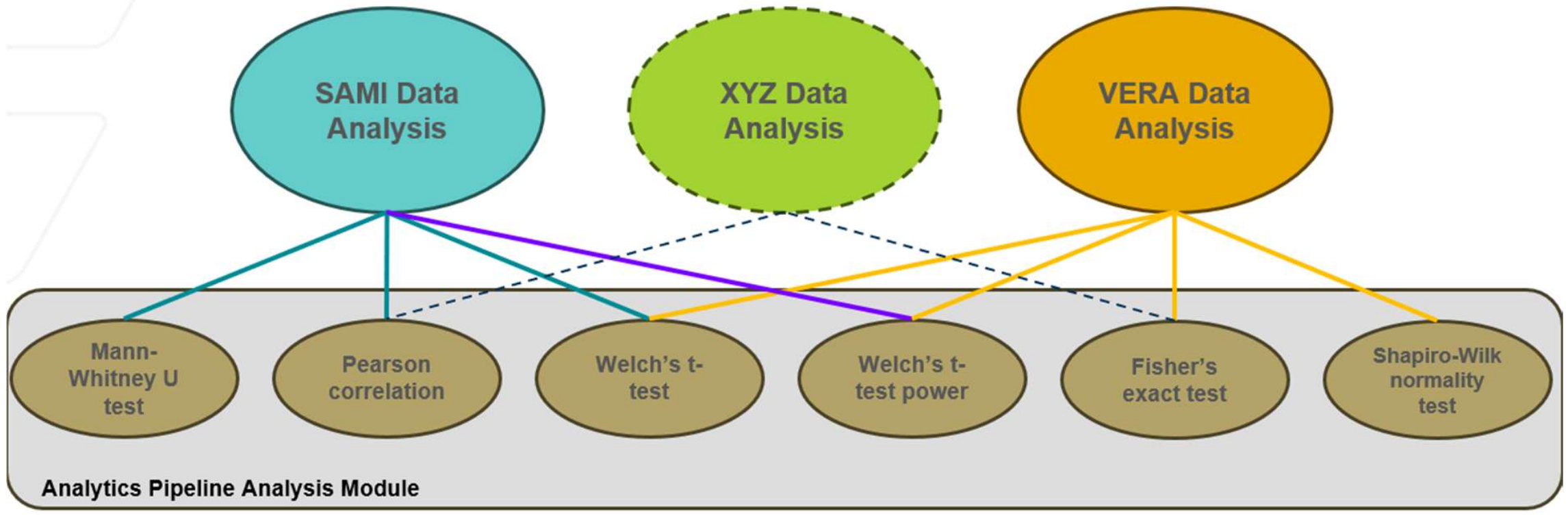

*Figure 3. Analytics Pipeline usage by different domains. Orange lines indicate which statistical options were used by the VERA analysis to replicate previous findings. Blue lines indicate which options were used by the SAMI analysis to replicate previous findings. The purple line indicates a statistical option not performed in the original SAMI analysis, but applicable to the research questions thanks to the extensibility and flexibility of the pipeline. An implementation of a hypothetical XYZ analysis is displayed in green with dashed lines suggesting potential use of existing statistical options in the analysis module to answer XYZ's research questions.*

## 6.1 Future Evaluation

Future research on the A4L Analytics Pipeline could explore how additional AI assistants in the A4L environment and their data interact with the pipeline. In addition to JW, VERA, and SAMI, A4L has additional education AI assistants that could take the place of XYZ in the future. One candidate could be Ivy [23], an AI coach promoting active learning when learners watch online videos in their courses. It is expected that key values in the analysis configuration will be modified to fit the AI assistant's domain, but we do not expect that the pipeline's codebase will need to be modified.

# 7 CONCLUSION

By observing the A4L Analytics Pipeline's support in answering research questions from three separate domains (JW, VERA, and SAMI) using one central and modularized platform, this research indicates that A4L's Data Analytics Pipeline is extensible and flexible in the context of real-world scenarios. This finding builds off previous work on educational data analytics pipelines which feature flexibility and generalizability but did not have documentation on any such real-world implementation [9], [10]. For a given statistical option, the Analytics Pipeline provides a unified access point that can be used by each domain thanks to the configurable inputs made available by the pipeline's modular design. Due to the Analytics Pipeline's generalizability, this research found that a statistical option originally added in the pipeline for one analysis could be extended to other analyses, enhancing insights gained from each additional analysis.

One limitation of this study was that the process of extending the Analytics Pipeline in a flexible manner was done by research team members who are familiar with the system's architecture.

This presents the opportunity to observe and evaluate how someone, possibly a researcher interested in another educational AI tool at Georgia Tech, incorporates functionality from their data analysis scripts into the pipeline. Observing how someone less familiar with the Analytics Pipeline extends its capabilities would provide further evidence to evaluate the pipeline as flexible platform that can accommodate a broader range of users. Furthermore, exploring this area would provide additional data to assess the generalizability of the pipeline's analysis functionality.

# ACKNOWLEDGEMENTS

We would like to thank members of the A4L team in Georgia Tech's Design Intelligence Laboratory for their guidance and encouragement during this process.